\documentclass{article}
\usepackage{graphicx}
\newcommand{\be}{\begin{equation}}
\newcommand{\ee}{\end{equation}}
\newcommand{\bea}{\begin{eqnarray}}
\newcommand{\eea}{\end{eqnarray}}

\newcommand{\eqref}[1]{(\ref{#1})}

\begin{document}

\title{Electric magnetic duality of the
confining\\
flux tube in Abelian-projected
gluodynamics\footnote{Talk given at 
``\uppercase{C}onfinement \uppercase{V}'',
\uppercase{G}argnano, 
\uppercase{I}taly, 10-14 \uppercase{S}ep. 2002.}
}

\author{Yoshiaki  Koma}

\date{\normalsize{\it{
Max-Planck-Institut f\"ur Physik, \\
F\"ohringer Ring 6, M\"unchen, Germany\\ 
E-mail: ykoma@mppmu.mpg.de}}}

\maketitle

\begin{abstract}
The structure of the 
Abelian-projected (AP) flux tube in SU(2) gauge theory
in the maximally Abelian gauge is studied by applying 
the Hodge decomposition to the Abelian Wilson loop.
It is shown that the profile of the AP flux tube 
has the same structure as the 
 classical flux tube solution in the
dual Abelian Higgs model, which
is composed of a Coulombic and a solenoidal 
electric field  linked by monopole supercurrent.
\end{abstract}

\vskip 3mm

\par
It is known that the profiles of the classical flux-tube 
solution in the dual Abelian Higgs (DAH) model and the 
Abelian projected (AP) flux tube, observed in 
lattice simulations within the Abelian projection scheme
in the maximally Abelian gauge (MAG), 
look quite 
similar~\cite{Singh:1993jj,Matsubara:1994nq,%
Bali:1998gz}.
We then ask how these two kinds of profiles
can be related each other.
This question would be non-trivial
since the AP flux tube contains the quantum effects
at work in non-Abelian lattice gauge simulations
while the DAH flux tube is a classical 
solution obtained
by solving the field equations. 
To answer the question, it is important 
to clarify the structure of both flux
tubes~\cite{Koma:2002uq,Koma}.

\par
The DAH model, after applying the Hodge
decomposition to the dual gauge field,
is written, in terms of its magnetic and
electric parts,  $B^{\rm mag}$
and $B^{\rm ele}= 2 \pi \Delta^{-1}\delta *\Sigma^{\rm  ele}$, 
as\footnote{We use differential form notations.}
\bea
S_{\rm DAH} 
&=&
\frac{\beta_{m}}{2} 
(d B^{\rm mag} +  2 \pi * C^{\rm ele} )^{2} \nonumber\\*
&&+
|(d +i(B^{\rm mag}+B^{\rm ele}) )\chi|^{2}
+ \lambda (|\chi|^{2}-v^{2})^{2} \; ,
\label{eq:DAH-action1}
\eea
where $*C^{\rm ele}= \Delta^{-1}\delta * j$
with an external electric current 
$j= - \delta \Sigma^{\rm ele}$, 
and $\chi$ denotes the complex scalar monopole field.
$\Sigma^{\rm ele}$ is the electric Dirac string.
By solving the field equations 
taking into account the boundary conditions
of fields determined by $B^{\rm ele}$, 
we see that the flux tube 
is led by the superposition of two 
components, a Coulombic 
electric field  $2 \pi * C^{\rm ele}$, directly 
induced by the electric charges,  
and a solenoidal electric field, $d B^{\rm mag}$,
linked by monopole 
supercurrent, $k \propto \delta d B^{\rm mag} \neq 0$.

\par
On the other hand, the leading action of 
AP gluodynamics (reconstructed from the
effective monopole action up to self-interaction 
term) with an external electric current $j$, 
after the Hodge decomposition on the 
AP gauge field, may be given by
\be
S_A = \frac{\beta_{e}}{2} 
(dA^{\rm ele}+ 2\pi *C^{\rm mag})^{2}
-i(A^{\rm ele}+ A^{\rm mag},j) \; ,
\label{eq:A_reg_action}
\ee
where $A^{\rm ele}$ and 
$A^{\rm mag} = 2 \pi \Delta^{-1}
\delta *\Sigma^{\rm mag}$
denote the electric and magnetic parts 
of the AP gauge field, and
$*C^{\rm mag} = \Delta^{-1}\delta * k$.
Here, $\Sigma^{\rm mag}$ is the magnetic Dirac string
and $k \equiv - \delta \Sigma^{\rm mag}$ 
the monopole current, which should be 
summed over by taking into account $\delta k=0$.
The electric coupling $\beta_{e}=4/e^{2}$ and the
magnetic coupling in the DAH model 
$\beta_{m}=1/g^{2}$ must
satisfy the Dirac quantization condition,
$4\pi^{2}\beta_{e}\beta_{m}=1$ $(eg=4\pi)$.
Here, the Abelian Wilson loop
consists of electric photon 
and magnetic monopole parts  as
$W_{A}[j] \equiv \exp [i(A^{\rm ele}+ A^{\rm mag},j)]
=W_{\mathit{Ph}}[j] \cdot W_{\mathit{Mo}}[j]$.
We may call 
$W_{\mathit{Ph}}[j]=\exp [i(A^{\rm ele},j)]$ 
and 
$W_{\mathit{Mo}}[j]=\exp [i(A^{\rm mag},j)]$ 
the photon Wilson loop and the 
monopole Wilson loop, respectively.

\par
The action (\ref{eq:A_reg_action}) is translated 
to the DAH action (\ref{eq:DAH-action1}) by
the the path-integral duality transformation.
Here, we find that 
while the photon Wilson loop is a source of 
$C^{\rm ele}$ in the DAH action,
the monopole Wilson loop provides $B^{\rm ele}$, 
through the relation
$(A^{\rm mag} ,j)+ (k, B^{\rm ele}) 
= - 2 \pi ( \Sigma^{\rm mag}, *\Sigma^{\rm ele}) 
= 2 \pi N$, where $N \in Z\!\!\! Z$.
From these duality relations we come to 
a conjecture that if there is a 
one-to-one correspondence
between the AP flux tube and the DAH 
flux tube as for the structure,
we should be able to observe
a composed structure for an AP flux tube
like a DAH flux tube
by measuring the profiles induced 
by the photon and the monopole Wilson 
loops:
while the photon Wilson loop 
induces a Coulombic electric field,
the monopole Wilson loop
leads to a solenoidal electric field
with monopole supercurrents.
The superposition then will lead
to a full AP flux 
tube.

\begin{figure}[!t]
\centering
\includegraphics[width=11.5cm]{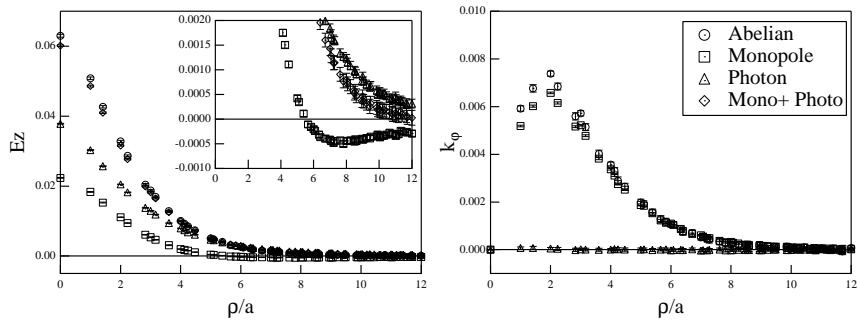}
\caption{
Profiles of the electric field 
(left) and the monopole current (right)
from correlators with 
Abelian, photon and monopole Wilson loops
at $r=6a=0.49$ fm.
}
\label{fig:fig1}
\end{figure}

\par
In Fig.~\ref{fig:fig1},
we show lattice results on
the flux-tube profiles (just on the midplane of 
the $q$-$\bar{q}$ system) 
measured with the complete 
Abelian, the photon, and the monopole Wilson 
loops as sources. 
The measurement has been done at $\beta=2.5115$
on a $32^{4}$ lattice after the MAG has been fixed.
The $q$-$\bar{q}$ distances are $r=6a$
with the lattice spacing $a = 0.081$ fm.
Inside the Fig.~\ref{fig:fig1} (left) we 
show the same electric field profiles, focussing
on the region where $E_{z} \sim 0$.
We see that these lattice results, 
the behavior of the profiles, 
strongly support our conjecture:
although the electric field from
the monopole Wilson loop
takes positive value near the center, 
it turns {\em negative} beyond a certain 
radius $\rho$ (in the given case, $\rho \sim 5.5~a$). 
This signals the appearance of a 
solenoidal electric field, which 
cancels the Coulombic field induced
by the photon Wilson loop.
The sum of these two contribution then
reproduces the profile from 
the complete Abelian Wilson loop.
There is no correlation between the 
photon Wilson loop and monopole current as 
expected.

\par
We then conclude that the structure of the AP flux tube 
and the DAH flux tube is basically the same.
This observation seems to support that 
the path-integral duality transformation, 
the mapping $e \to g$, works
very well and then allows us indeed to have
an infrared effective theory which can be
treated perturbatively or even in a classical manner.
Behind this idea, in particular, 
we find that the singularities associated
with the magnetic Dirac and the electric Dirac strings
play an important role to connect both theories.
This connection allows us to see the 
similar flux-tube profile
no matter whether in the AP gauge theory 
or in the DAH model.

\par
The author is grateful to
M.~Koma, E.-M. Ilgenfritz, T.~Suzuki and 
M.I.~Polikarpov for fruitful collaboration.
The computation was done on the Vector-Parallel Supercomputer 
NEC SX-5 at the RCNP, Osaka University, Japan.

\end{document}